\documentclass{ceab}   

\usepackage{epsfig}     
\usepackage{graphicx}   

\usepackage{ceabbib}     
\usepackage[T1]{fontenc}

\begin{document}

\def\tit{A Poorly Studied Binary BG~Gem}
\def\aut{Drake et al.}

\title{BG Gem -- a Poorly-Studied Binary with a Possible Black Hole Component}

\author{N.~A. Drake$^{1,2}$, A.~S. Miroshnichenko$^3$, S. Danford$^3$,
and C.~B. Pereira$^1$\\
\it $^1$ Observat\'orio Nacional/MCTI, Rio de Janeiro, Brazil\\
\it $^2$ Sobolev Astronomical Institute, Saint-Petersburg State University,\\
\it Saint-Petersburg, Russia\\
\it $^3$University of North Carolina at Greensboro,\\
\it Department of Physics and Astronomy, Greensboro, NC, USA\\}

\maketitle

\begin{abstract}
BG~Gem is an eclipsing binary with a 91.6--day orbital period. The
more massive primary component does not seem to show absorption
lines in the spectrum, while the less massive secondary is thought
to be a K-type star, possibly a supergiant. These results were
obtained with optical low-resolution spectroscopy and photometry.
The primary was suggested to be a black hole, although with a low
confidence. We present a high-resolution optical spectrum of the
system along with new $BVR$--photometry. Analysis of the spectrum
shows that the K-type star rotates rapidly at $v\,\sin i =
18$~km\,s$^{-1}$ compared to most evolved stars of this temperature
range. We also discuss constraints on the secondary's luminosity
using spectroscopic criteria and on the entire system parameters
using both the spectrum and photometry.
\end{abstract}


\section{Introduction}

BG~Gem is a 13--mag stellar object that shows brightness variations
with a period of 91.645 days and an amplitude of $\sim$ 0.5 mag at
visible wavelengths (Benson et al. 2000). The primary eclipse is
only observed at $\lambda \le$ 4400\,\AA, while the light curves are
shallower at longer wavelengths. From low-resolution ($\sim$6 \AA)
observations of the system, Kenyon, Groot, \& Benson (2002) deduced
that the optical spectrum is dominated by absorption lines from a
K0~{\sc i} secondary component. They suggested that line emission
which occurs in the Balmer and some He\,{\sc i} lines may originate
from a B--type or a black hole primary component. From the radial
velocity curve, they determined a components mass ratio of
0.22$\pm$0.07.

Elebert et al. (2007) took a high-resolution optical spectrum,
retrieved an archival UV spectrum obtained with HST, and found an
upper limit for the X-ray flux from INTEGRAL. They detected a weak
UV continuum (more indicative of a black hole than of a normal
star), but a low X-ray luminosity ($\le 10^{35}$ erg\,s$^{-1}$) and
a low rotational velocity of the disk material (500 km\,s$^{-1}$ in
the Balmer lines and $\le$ 1000 km\,s$^{-1}$ in the UV emission
lines). Therefore, these findings are inconclusive. However, if
BG~Gem does have a black hole component, it becomes the
longest-period system of this kind in the Milky Way.

Open questions in studies of this system include the poorly known
fundamental parameters of the cool secondary companion, the unknown
nature of the primary component, and the evolutionary status of the
entire system.

\section{Observations}

In order to refine the secondary component parameters as well as
those of the entire system and check the light curve, we took new
spectroscopic and photometric observations of the BG~Gem system.

\begin{figure}[htb]
\begin{center}
\includegraphics[width=7.5cm,height=7cm]{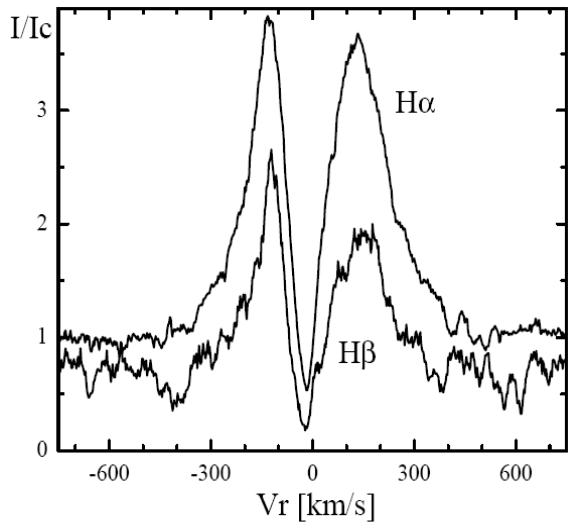}
\end{center}
\caption{Balmer line profiles in our spectrum of BG~Gem. The
intensity is normalized to the nearby continuum and plotted against
the heliocentric radial velocity.} \label{f1}
\end{figure}

The photometric data were obtained in the $VRH\alpha$ filters during
11~nights in 2005--2006 at the 0.81--m telescope of the Three
College Observatory near Greensboro, North Carolina, USA. The main
goal of these observations was to look for changes of the light
curves compared to the published data. {\bf Our data show the same
brightness level at covered phases, but they are insufficient to
detect any possible change of the orbital period.}

\begin{figure}[htb]
\begin{center}
\includegraphics[width=10cm]{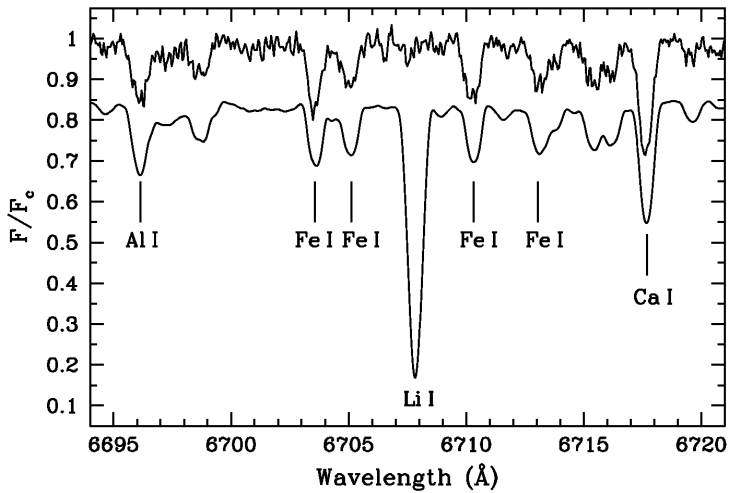}
\end{center}
\caption{Comparison of our spectrum of BG~Gem (upper line) and that
of a K1~{\sc iii} star HD~19745 obtained with the same resolution.
The spectral lines of HD~19745 were broadened to a rotational
velocity of 18~km\,s$^{-1}$ to match the line widths of BG~Gem. The
spectrum of BG~Gem resembles the spectrum of HD~19745 except for the
Li\,{\sc i} line at $\lambda6707.8$~\AA\ which is absent in the
spectrum of BG~Gem.} \label{f2}
\end{figure}

A high--resolution \'echelle spectrum ($\lambda\lambda$
4800--10500~\AA, $R$ = 60\,000) was obtained at the 2.7-m Harlan J.
Smith telescope of the McDonald Observatory in December 2006. The
spectrum was taken at an orbital phase of 0.62 (from the primary
minimum). It has a signal-to-noise ratio of $\sim$30 in continuum
near $\lambda$4800 \AA\ that rises towards longer wavelengths
reaching $\sim$100 near $\lambda$\,1 $\mu$m. Parts of the spectrum
are shown in Figures \ref{f1} and \ref{f2}.

\begin{figure}[htb]
\begin{center}
\begin{tabular}{cc}
\includegraphics[width=6cm,height=5cm]{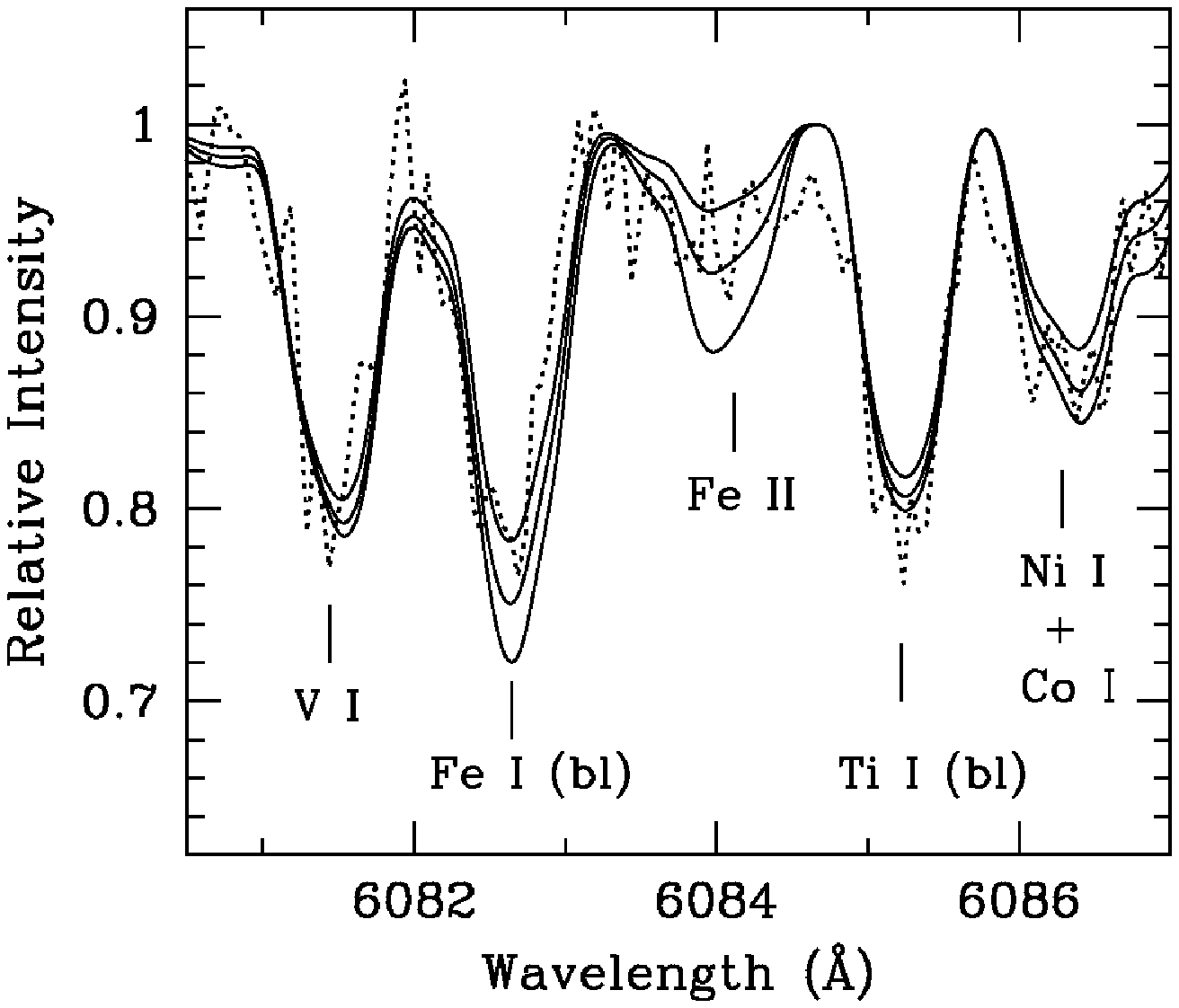}&
\includegraphics[width=6cm,height=5cm]{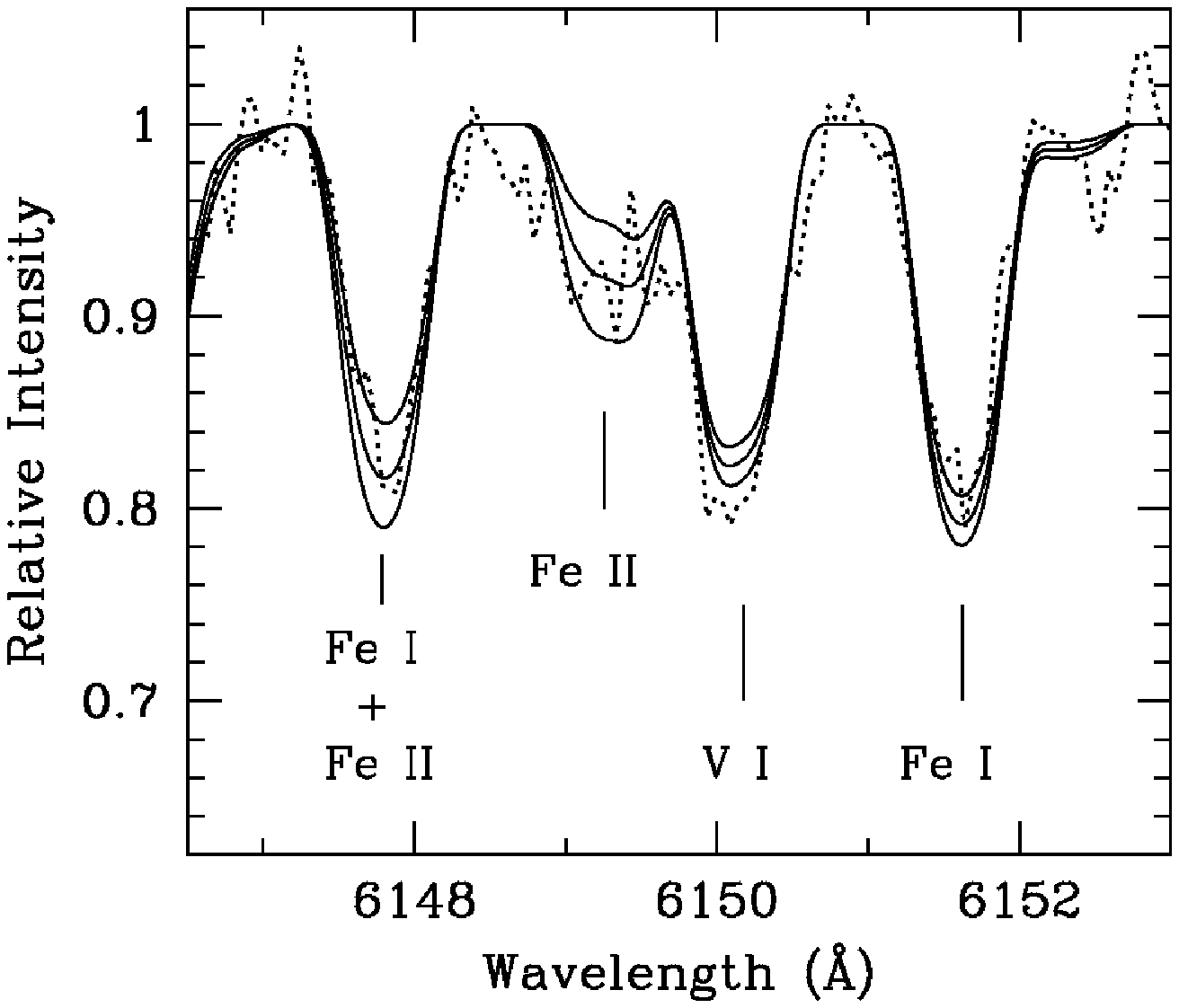}\\
\end{tabular}
\end{center}
\caption{Comparison of the observed BG~Gem spectrum (dotted line)
with synthetic spectra calculated for $T_{\rm eff}$ = 4500~K and
$\log g$ = 0.8, 1.6, and 2.4 (from bottom to top).} \label{f3}
\end{figure}

\section{Atmospheric parameters}

Significant rotational broadening of the lines in the spectrum of
BG~Gem, an insufficient S/N ratio, and veiling of the spectrum
caused by the primary's accretion disk emission make it difficult to
measure equivalent widths of individual Fe\,{\sc i} and Fe\,{\sc ii}
lines and perform the usual analysis based on excitation and
ionization equilibrium.

To derive the atmospheric parameters of BG~Gem, we compared its
spectrum with those of various K--type giant and supergiant stars
with well-known parameters. We broadened lines in these spectra to
the rotation velocity of BG~Gem.

Projected rotation velocity of BG~Gem was determined by means of
fitting the observed spectrum with synthetic spectra, calculated
with different values of rotation velocity. Taking into account the
FWHM of the instrumental profile of the spectrograph and adopting a
macroturbulent velocity of 2~km\,s$^{-1}$, we determined a rotation
velocity of $v\sin i =18.0\pm 1.0$~km\,s$^{-1}$ for BG~Gem.

Using these parameters, we calculated a grid of synthetic spectra in
a wide range of effective temperatures and surface gravities
($T_{\rm eff}/\log g$) in the spectral region $6080 - 6180$~\AA\
containing lines of neutral elements with different excitation
potentials and two lines of Fe\,{\sc ii} (at $\lambda 6084.1$~\AA\
and $\lambda 6149.2$~\AA) which are sensitive to surface gravity.

Synthetic spectra were calculated using the local thermodynamic
equilibrium (LTE) model atmospheres of Kurucz (1993) and the current
version (August 2010) of the spectral analysis code {\sc moog}
(Sneden 1973). The solar abundances taken from Anders \& Grevesse
(1989) were adopted except the carbon and nitrogen abundances, which
are known to be modified in giant stars (see, e.g., Lambert \& Ries
1981). The VALD atomic data base (Kupka et al. 1999) was used to
create the line list. A microturbulent velocity of 2.0~km\,s$^{-1}$
was adopted in our synthetic spectra calculations. An additional
flux of 30\% in continuum ($F_{\rm disk}/F_{\rm cont}$) was added to
the synthetic spectra.

As an example, in Fig. \ref{f3} we show a comparison of the observed
and synthetic spectra calculated with $T_{\rm eff} = 4500$~K and
different values of $\log g$ (from 0.8 to 2.4) in two spectral
regions containing Fe\,{\sc ii} lines at $\lambda 6084.1$~\AA\ and
$\lambda 6149.2$~\AA\ used to estimate the surface gravity.

Among high-rotating K giants with infrared excesses, nearly half
have a high Li abundance (Drake et al. 2002). Such an abundance was
found by Gonz\'alez Hern\'andez et al. (2004) for the secondary
component in the black hole binary A0620--00 and by Sabbi et al.
(2003) in the companion star to the millisecond pulsar J1740--5340
in NGC~6397 suggesting some Li production in these systems. We
calculated synthetic spectra in the  region of the Li\,{\sc i}
6708~\AA\ resonance line and found a low Li abundance in BG~Gem,
$\log\varepsilon{\rm (Li)} \le 0.2$ (in the notation
$\log\varepsilon{\rm (X)}=\log(N_{\rm X}/N_{\rm H}) +12.0)$.

\section{Results}

Analysis of the absorption line spectrum shows that the secondary
component is a K2~{\sc ii}/{\sc iii} star with a projected
rotational velocity of 18$\pm 1$~km\,s$^{-1}$, an effective
temperature of 4500$\pm$300~K, and a surface gravity $\log g$ =
1.5$\pm$0.5. The derived surface gravity implies a mass of the
secondary component of 1\,$M_\odot$, if it is filling its Roche
lobe. Diffuse interstellar bands are weak implying a noticeable
contribution of the primary's disk to the near-IR excess. Assuming
an interstellar extinction of $A_V$ = 1.5~mag and $M_{V}\!=\!
-0.8$~mag for the K2--giant, the system is located at a distance of
at least 3.5~kpc (no disk contribution at its eclipse by the
secondary). No features that could be a result of an explosion in
the system due to formation of the black hole are seen in the {\it
WISE} data between $\lambda$3.4 and $\lambda$11.5 $\mu$m.

\section{Conclusions}

We refined the fundamental parameters of the visible star in the
BG~Gem binary system and showed that it is a rapidly rotating
K2~{\sc ii}/{\sc iii} star rather than a K0~{\sc i} star suggested
by Benson et al. (2000) and Kenyon et al. (2002).

It is still unclear whether a black hole is present in the system.
The fact that we detect no light from a hot star implies that the
primary's disk completely veils it. Spectroscopic observations at
both eclipses are needed to put further constraints on the system
properties.

\section*{Acknowledgements}
A.M. acknowledges support from the American Astronomical Society
International Travel Grant program and from the Department of
Physics and Astronomy of the University of North Carolina at
Greensboro. N.A.D. acknowledges support of a PCI/MCTI (Brazil) grant
under the Project 311.868/2011--8. This paper is party based on
observations obtained at the McDonald Observatory of the University
of Texas at Austin. This research has made use of the SIMBAD
database, operated at CDS, Strasbourg, France, as well of the
Wide-Field Infrared Survey Explorer ({\it WISE}) database.

\section*{References}
\begin{itemize}
\small
\itemsep -2pt
\itemindent -20pt
\item[] Anders, E., and Grevesse, N.: 1989, {\it Geochim. et Cosmochim. Acta}, {\bf 53}, 197.
\item[] Benson, P., Dullighan, A., Bonanos, A., McLeod, K.K., and Kenyon, S.J.: 2000, {\it \aj} {\bf 119}, 890.
\item[] Drake, N.A., de la Reza, R., da Silva, L., and Lambert, D. L.:
        2002, {\it \aj} {\bf 123}, 2703.
\item[] Elebert, P., Callanan, P.J., Russel, L., and Shaw, S.E.:
        2007, {\it Proc. IAU Symp.} {\bf 238}, 361.
\item[] Gonz\'alez Hern\'andez, Rebolo, R., Israelian, G., and Casares, J.: 2004,
        \apj, {\bf 609}, 998.
\item[] Kenyon, S.J., Groot, P.J., and Benson, P.: 2002, {\it \aj} {\bf 124}, 1054.
\item[] Kupka, F., Piskunov, N., Ryabchikova, T.A., Stempels, H. C., and Weiss, W.W.:
        1999, {\it \aaps} {\bf 138}, 119.
\item[] Kurucz, R.L.: 1993, {\it CD-ROM No. 13, Smithsonian Astrophysical Observatory}.
\item[] Lambert, D.L., and Ries, L.M.: 1981, {\it \apj} {\bf 248}, 228.
\item[] Sabbi, E., Gratton, R.G., Bragaglia, A., Ferraro, F.R.,
        Possenti, A., Camilo, F., and D'Amico, N. : 2003, {\it \aap} {\bf 412}, 829.
\item[] Sneden, C.: 1973,  Ph.D. Thesis, Univ. of Texas.
\end{itemize}

\end{document}